%% file: main.tex
\documentclass[aps,prd,twocolumn,showpacs,superscriptaddress,groupedaddress]{revtex4-1} 
\usepackage{graphicx}  
\usepackage{dcolumn}   
\usepackage{bm}        
\usepackage{amssymb}   

\usepackage{lineno} 
\usepackage{booktabs}
\usepackage{longtable}
\usepackage{overpic}
\usepackage{multirow}
\usepackage{amsmath}
\usepackage{color}
\usepackage{xspace}

\newsavebox{\tablebox}

\def\BES {\text{BES\uppercase\expandafter{\romannumeral3}}}
\def\BTWO {\text{Belle{\ }\uppercase\expandafter{\romannumeral2}}}

\def\massRes{\ensuremath{2239.2\pm7.1~{\rm MeV}/c^2}\xspace}
\def\massResys{\ensuremath{2239.2\pm7.1\pm11.3~{\rm MeV}/c^2}\xspace}
\def\widthRes{\ensuremath{139.8\pm12.3~{\rm MeV}}\xspace}
\def\widthResys{\ensuremath{139.8\pm12.3\pm20.6~{\rm MeV}}\xspace}

\begin{document}


\title{\boldmath Measurement of $e^{+} e^{-} \rightarrow K^{+} K^{-}$ cross section at $\sqrt{s} = 2.00 - 3.08$ GeV}
\input{authors}

\date{\today}

\begin{abstract}
The cross section of the process $e^{+} e^{-} \rightarrow K^{+} K^{-}$ is measured at a number of center-of-mass energies $\sqrt{s}$ from 2.00 to 3.08 GeV with the BESIII detector at the Beijing Electron Positron Collider (BEPCII). 
The results provide the best precision achieved so far.
A resonant structure around 2.2 GeV is observed in the cross section line shape.
A Breit-Wigner fit yields 
a mass of $M=\massResys$~and a width of $\Gamma=$~\widthResys, where the first uncertainties are statistical and the second ones are systematic.
In addition, the time-like electromagnetic form factor of the kaon is
determined at the individual center-of-mass energy points.
\end{abstract}

\maketitle

\section{Introduction}
\indent

The study of the hadron spectrum provides important input to
understand the non-perturbative behavior of QCD.  In the full hadron
spectrum, the spectrum of light mesons has a particular position since there exist abundant data on light mesons.
However, a further check of the experimental data on the light mesons listed in Particle Data Group (PDG) \cite{Tanabashi:2018oca} reveals that many light mesons with a mass above 2 GeV are far from being firmly established.  This poses a challenging task to the experimentalist community.

In the past years, experimentalists have spent considerable effort on this issue. A typical example is $Y(2175)$ observed by the {BaBar}
Collaboration in 2006 in the process $e^+e^-\to \gamma_{ISR} \phi f_0(980)$ \cite{Aubert:2006bu}, 
which was confirmed by the Belle, BESII, and BESIII experiments\cite{Aubert:2007ur,Aubert:2007ym,Ablikim:2007ab,Shen:2009zze,Lees:2011zi,Ablikim:2014pfc}.  
The discovery of the $Y(2175)$ has stimulated extensive discussion about its internal structure;
proposed solutions include an $s\bar{s}g$ hybrid state \cite{Ding:2006ya}, $3S$ \cite{Barnes:2002mu} and $2D$ \cite{Wang:2012wa,Ding:2007pc} states in the conventional $\phi$ family, $s\bar{s}s\bar{s}$ tetraquark state \cite{Wang:2006ri,Chen:2008ej}, $\Lambda\bar{\Lambda}$ baryonium \cite{Klempt:2007cp}, 
$\phi f_0(980)$ resonance \cite{MartinezTorres:2008gy} and $s$-quark counterpart to the $Y(4260)$ \cite{Yuan:2007sj}. 
Although the $Y(2175)$ is now denoted as $\phi(2170)$ by the Particle Data Group (PDG) \cite{Tanabashi:2018oca}, its properties still need to be clarified by further theoretical and experimental effort.  Under different hypotheses for the internal structure, the $Y(2175)$ can have common decay channels but with different decay rates, such as the decay $Y(2175)\to K \bar{K}$  \cite{Ding:2006ya,Barnes:2002mu,Wang:2012wa,Ding:2007pc}.  
In the flux tube and $^{3}P_{0}$ models, when treating $Y(2175)$ as a $s\bar{s}g$ or $3^{3}S_{1}s\bar{s}$ state, the ratio of the partial width of the $K \bar{K}$ channel to the total width is predicted to be almost zero compared to other channels, while the $2^{3}D_{1}$ state hypothesis predicts a branching fraction of about $5-10$\% \cite{Ding:2007pc}. 
This provides a powerful tool to distinguish between models, and  
a more precise measurement of $e^+e^-\to K\bar{K}$ using BESIII data is highly desirable.

Much effort has been spent to understand the process $e^{+} e^{-} \rightarrow K^{+} K^{-}$~\cite{DM2KK,CMD2KK, BABARKK,BABARKK2,eeSUM,BerKK,CLEOKK}.
Previous experiments have achieved cross section uncertainties of a few percent in the energy region around the $\phi(1020)$, while above 2.0~GeV, the uncertainties are larger than 15\%.
The {BaBar} collaboration measured the $e^{+} e^{-} \rightarrow K^{+} K^{-}$ cross section using the Initial State Radiation (ISR) technique. Their measurements range from the $K^{+} K^{-}$ threshold up to 8 GeV, and some complicated structures between 1.8 and 2.4 GeV \cite{BABARKK,BABARKK2} are observed.
In this paper, we measure the $e^{+} e^{-} \rightarrow K^{+} K^{-}$ process directly using data collected in an energy scan at 22 energies from 2.00 to 3.08~GeV. The individual luminosities of each data point range from 1 to 126 pb$^{-1}$.

Besides the $Y(2175)$, there exist higher excitations of the $\rho$ and $\omega$ meson families located in the same mass range \cite{Clegg:1989mp,Biagini:1990ze,Aubert:2007ef,Lees:2012cj,Anisovich:2011sva}. For example, $\rho(2150)$ was reported by {BaBar} in the process $e^+e^-\to (\gamma) \pi \pi$~\cite{Lees:2012cj}. 
These reported or predicted higher excitations of $\rho$ and $\omega$ may also decay into $K\bar{K}$ \cite{He:2013ttg,Wang:2012wa}. 
Thus, measuring the process $e^{+} e^{-} \rightarrow K^{+} K^{-}$ can provide important information on these higher excitations of the $\rho$ and $\omega$ meson families around 2 GeV, which is crucial to construct the $\rho$ and $\omega$ meson spectra.

Additionally, in this work we report measurements of the kaon form factor $F_{K}(Q^2)$ through the obtained $e^+e^-\to K\bar{K}$ data.
The structure of light hadrons, parameterized in terms of electromagnetic form factors, is crucial to understand the internal dynamics of hadrons, the detailed structure of hadronic wave functions, and the nuclear and hypernuclear forces~\cite{HadFF, EMStructure}.
The form factor can be split into two categories, spacelike (momentum transfer $Q^{2}>0$) and timelike ($Q^{2}<0$) form factors. Spacelike form factors are directly associated with the charge distribution in hadrons, which are difficult to measure at large momentum transfers, and can only be obtained by analytic continuation of timelike form factors.
Precision measurements of timelike form factors at the highest possible momentum transfers are needed.
Perturbative QCD (pQCD) predicts the kaon form factor $F_{K}(Q^2)$ asymptotically to be inversely proportional to the center-of-mass energy; this can be tested by a precise measurement of $F_K$.

\section{Detector and Data Samples}
BEPCII \cite{ACC, BEPCDES} is a double-ring $e^{+}e^{-}$ collider optimized for a luminosity of $10^{33}$~cm$^{-2}$s$^{-1}$ at $\sqrt{s} = 3.770$~GeV. 
The BESIII detector \cite{ACC,Detector} is located at the collision point of BEPCII and has a geometrical acceptance of 93\% of the full solid angle. BESIII has five main components: 
i) A small-cell, helium-based (60\% He, 40\% C$_{3}$H$_{8}$) main drift chamber (MDC) with 43 layers providing an average single-hit resolution of 135 $\rm \mu m$ and a momentum resolution in a 1~T magnetic field of 0.5\% at 1~GeV/$c$; 
ii) A time-of-flight (TOF) system used for particle identification. It is composed of 5~cm thick plastic scintillators, with 176 detectors of 2.4~m length in two layers in the barrel and 96 fan-shaped detectors in the endcaps.
The barrel (endcap) time resolution of 80~ps (110~ps) provides $2\sigma K/\pi$ separation for momenta up to 1.0~GeV/$c$;
iii) A cylindrical electromagnetic calorimeter (EMC) consisting of a barrel and two endcaps. 
The energy resolution for electrons or photons with 1.0~GeV energy is 2.5\% (5\%) in the barrel (endcaps), and the position resolution is 6~mm (9~mm), respectively; 
iv) A super-conducting magnet generating a 1~T magnetic field at a current of 3400~A; 
v) A muon system (MUC) in the iron flux-return yoke of the magnet, consisting of 1272 m$^{2}$ of resistive plate chambers (RPCs) in nine barrel and eight endcap layers, providing $2$ cm position resolution.

The data samples used in this analysis were collected with the BESIII detector at 22 center-of-mass (c.m.) energies between  2.00 and 3.08~GeV
and correspond to a total integrated luminosity of 651~pb$^{-1}$~\cite{GaoZhen}.
Monte Carlo (MC) simulated samples of signal and background processes are used to optimize the event selection criteria, evaluate the reconstruction efficiency and estimate the background contamination.
The signal MC sample of $e^{+} e^{-} \rightarrow K^{+} K^{-}$ was generated using the package~\textsc{conexc} \cite{CONEXC}, which incorporates the radiative correction factors for the higher-order process with one photon in the final state.
Background samples of the processes $e^{+} e^{-} \rightarrow e^{+} e^{-}$, $\mu^{+} \mu^{-}$ and $\gamma \gamma$ are generated with the \textsc{babayaga}~\cite{BABAYAGA} generator, while the \textsc{luarlw}~\cite{LUARLW} and \textsc{bestwogam}~\cite{BESTWOGAM} generators are used for other background channels, including the processes $e^{+} e^{-} \rightarrow \textrm{hadrons}$ and $e^{+} e^{-} \rightarrow e^{+} e^{-} X$ (where $X$ denotes hadrons or leptons).

The generated particles are propagated through a virtual detector using a \textsc{geant\footnotesize4}-based~\cite{GEANT4} simulation software package \textsc{BESIII Object Oriented Simulation Tool} \cite{BOOST}, which includes the description of geometry and materials, particle transport and detector response. The MC simulation are digitized and tuned to experimental running conditions.

\section{Event selection}
\begin{figure}[b]
   \centering 
   \begin{overpic}[scale=0.4]{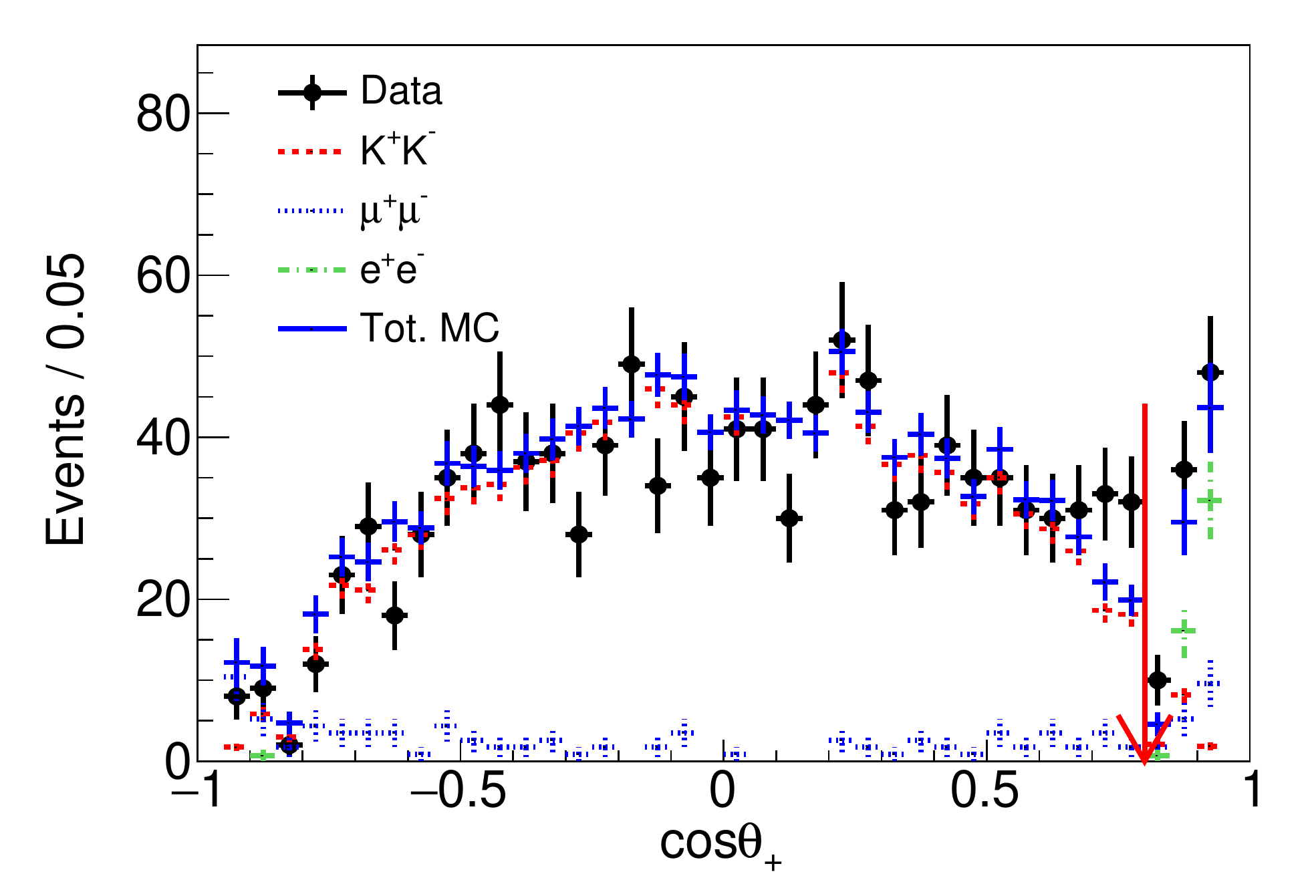}
   \put(60,55){} \end{overpic}
   \begin{overpic}[scale=0.4]{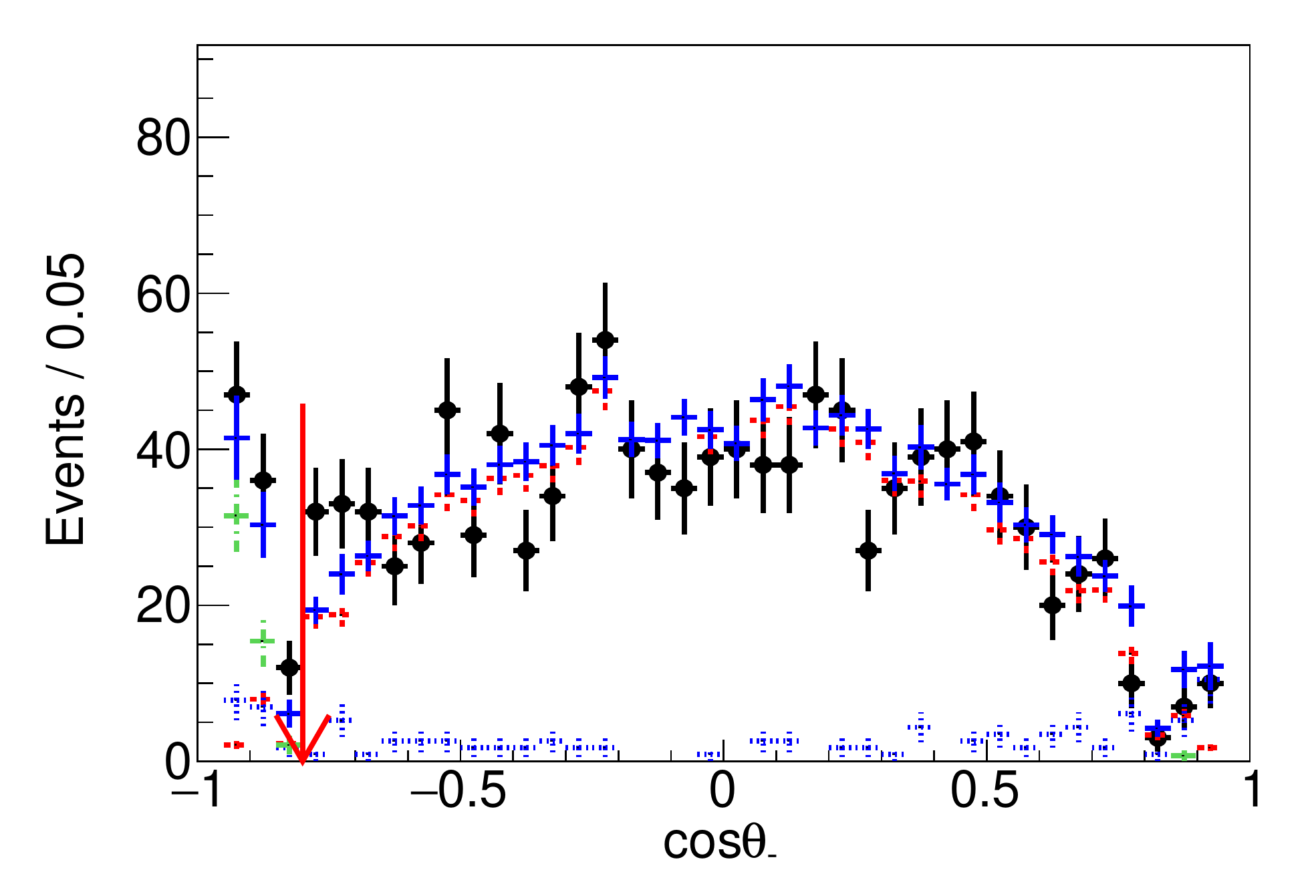}
   \put(60,55){} \end{overpic}
   \caption{(color online) Polar angle distribution of positive (upper) and negative (lower) tracks at $\sqrt{s} = 2.6444$~GeV after performing all selection criteria, as well as the requirement of the momenta of both tracks to be within the region of 3 times of resolution except for the $\cos\theta$ requirements. The arrows show the corresponding requirements on the polar angle distribution. "Tot. MC" in the legend means the sum of signal (red-dashed lines) and the dominant backgrounds, $e^{+} e^{-} \rightarrow (\gamma) \mu^{+} \mu^{-}$ (blue dotted) and $e^{+} e^{-} \rightarrow (\gamma) e^{+} e^{-}$ (green dot-dashed), estimated by MC simulation.
}
   \label{PolAng}
\end{figure}

\begin{figure}[t]
   \centering
   \includegraphics[scale=0.4]{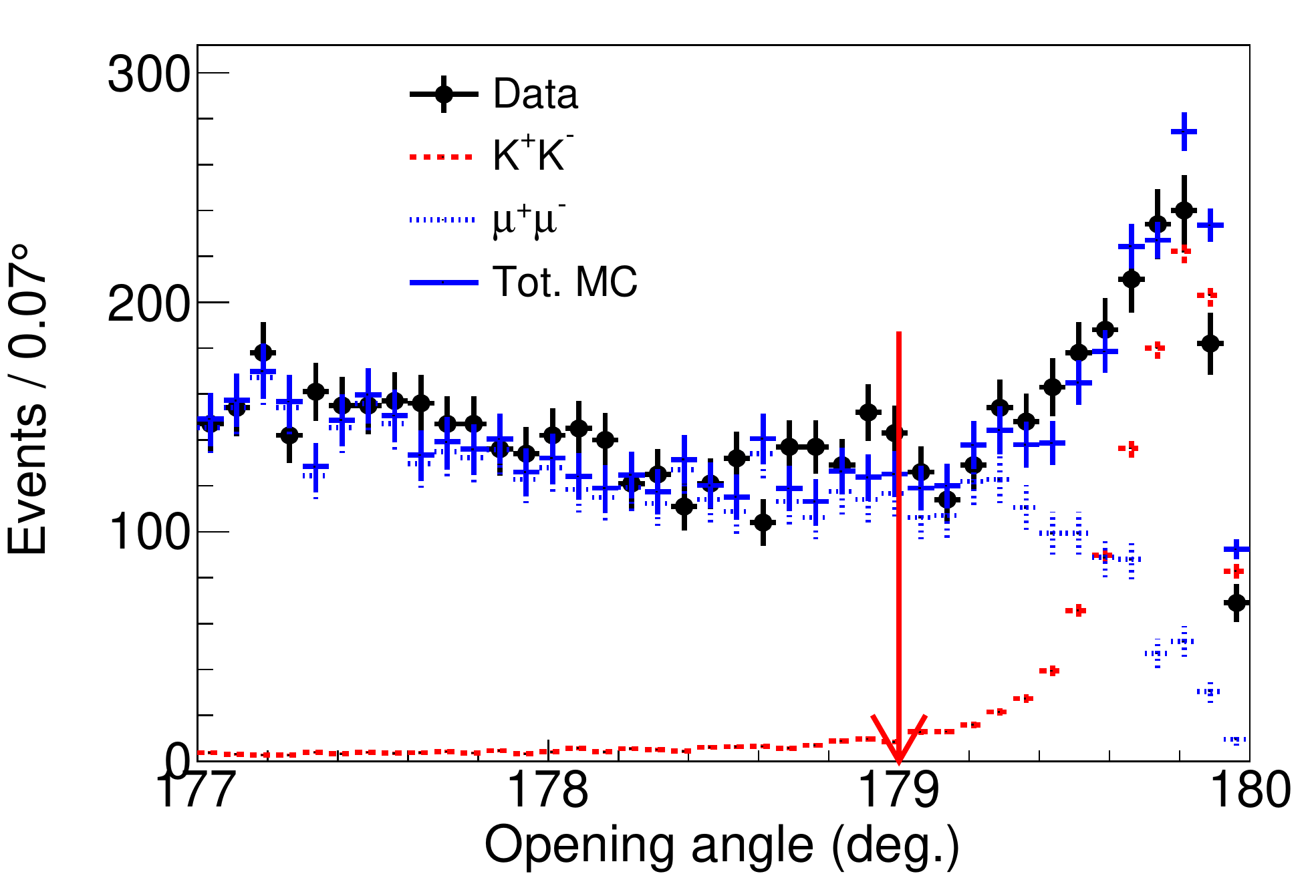} 
   \caption{(color online) Opening angle between the two charged tracks at $\sqrt{s} = 2.6444$ GeV after performing all selection criteria, as well as a requirement on the momenta of the negative track be within the region of 3 times of resolution except for the opening angle requirement. The arrow shows the corresponding selection requirement.
 "Tot. MC" in the legend means the sum of signal (red-dashed lines) and the backgrounds $e^{+} e^{-} \rightarrow (\gamma) \mu^{+} \mu^{-}$ (blue dotted).
   }
   \label{OpenAng}
\end{figure}

The signal candidates are required to have two oppositely charged tracks within the MDC coverage, $|\cos\theta| < 0.93$, where $\theta$ is the polar angle of the charged track. Each charged track is required to originate from a cylinder around the interaction point of 1 cm radius and extending $\pm10$~cm along the beam direction.
To suppress background of $e^{+} e^{-} \rightarrow (\gamma) e^{+} e^{-}$, two criteria are implemented, \emph{viz.}, each charged track must have the ratio  $E/p$ of the energy measured in the EMC ($E$) to the momentum measured in the MDC ($p$) smaller than a certain value ranging between 0.7 and 0.8, where the chosen value depends on the c.m.\ energy and is optimized by maximizing the ratio of signal to background;
additionally, $\cos\theta < 0.8$ is required for the positive charged track, and $\cos\theta> -0.8$ for the negative charged track.
To suppress the background events with a multi-body final state, the opening angle between the two charged tracks in the $e^{+}e^{-}$ c.m.\ system is required to be larger than $179^{\circ}$.
To reject background from cosmic rays, the difference of time of flight between the two charged tracks, as measured by the TOF system, is required to be less than 3~ns.
Comparisons of the distributions of polar angular and the opening angle for the candidate events between data and MC simulation at c.m.\ energy $\sqrt{s} = 2.6444$~GeV are depicted in Figs.~\ref{PolAng} and \ref{OpenAng}, respectively, where good agreement is observed.

\begin{figure}[htbp]
	\centering
	\includegraphics[scale=0.42]{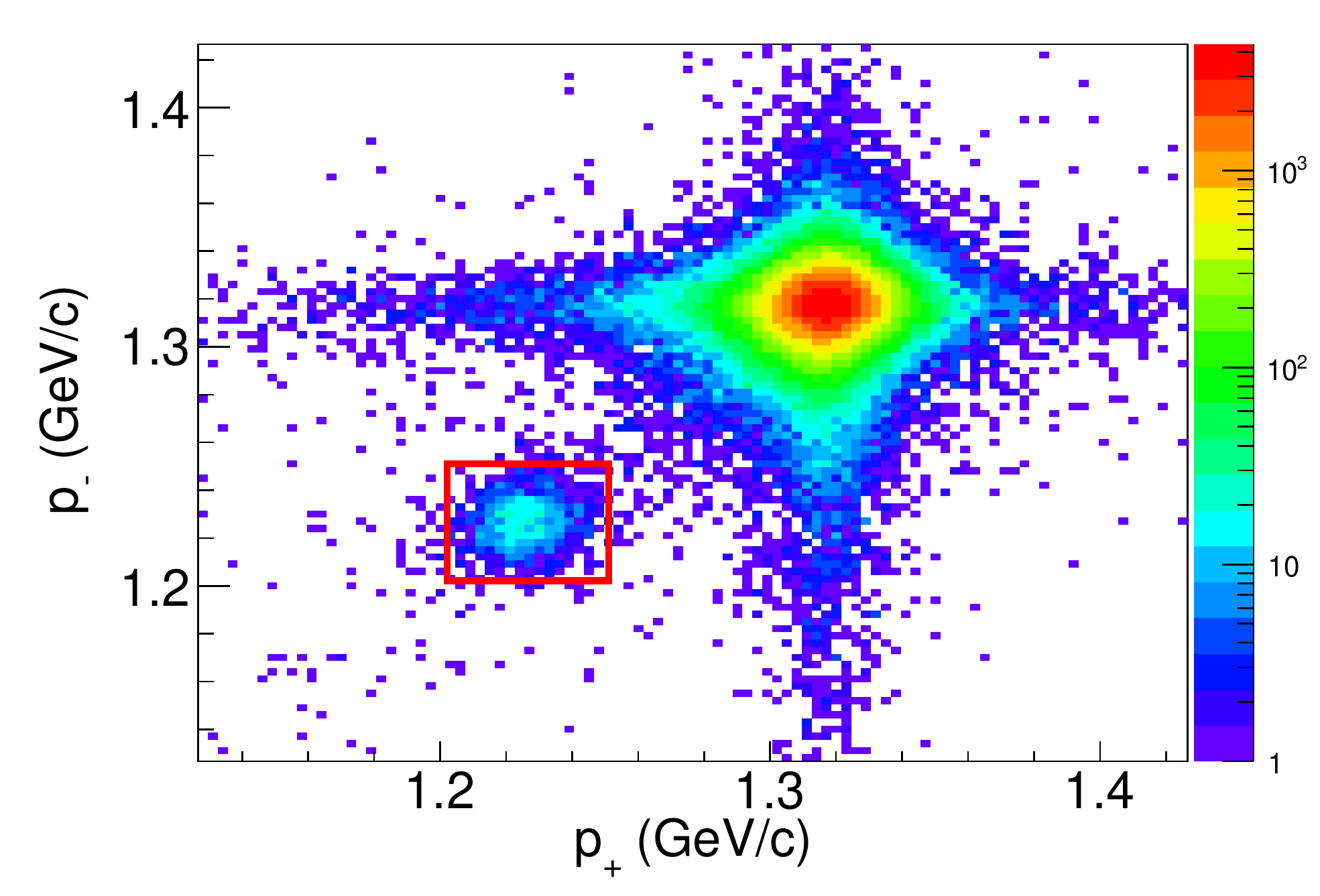}
	\caption {Scatter plot of the momentum of the positive track ($p_{+}$) versus that of the negative track ($p_{-}$) at $\sqrt{s} = 2.6444$ GeV. The signal events (3$\sigma_{p}$ region as shown in box) are concentrated around $p_{\pm} = 1.23\ {\rm GeV}/c$, while the $e^{+}e^{-}\rightarrow (\gamma)\mu^{+} \mu^{-}$ background accumulates around $p_{\pm} = 1.32$ GeV/$c$.
	}
	\label{2Dp}
\end{figure}

Since the process of interest is a two-body final state, the momenta of the charged tracks fulfil $p_{\textrm{exp}} = \sqrt{s/4 - m_{K}^{2}c^{4}}/c$, where $m_{K}$ is the $K^{\pm}$ mass.  This enables an efficient separation of the signal from background.  The momenta of positive charged tracks versus that of negative charged tracks of candidate events is illustrated in Fig.~\ref{2Dp}, where two clusters of events are observed, corresponding to the signal candidates (around $p_{\pm} = 1.23$~GeV/$c$) and background from $e^{+}e^{-}\rightarrow (\gamma)\mu^{+} \mu^{-}$ (around $p_{\pm} = 1.32$ GeV/$c$), respectively.

\section{Background analysis}
Potential sources of background are hadronic processes with multi-body final states and $e^{+} e^{-}$ annihilation into two-body final states, \emph{e.g.}, $e^{+} e^{-}$, $\mu^{+} \mu^{-}$ and $\pi^{+} \pi^{-}$, in which radiative processes reduce the momenta of the final-state particles so that they fall in the momentum region of kaons.
The level of background contamination is evaluated by MC simulations, with the momentum within a window of 3$\sigma_{p}$ around the signal, where $\sigma_{p}$ is the momentum resolution, 8~MeV/c at $\sqrt{s}=2.6444$~GeV.
The equivalent luminosities of the MC samples are between one to tens times of data for the different processes, individually, depending on the size of samples.
The backgrounds are found to be negligible for the processes $e^{+} e^{-} \rightarrow (\gamma) e^{+} e^{-}$, $\gamma \gamma$, and $e^{+} e^{-} X$, while they are estimated to be less than 0.5\% for the process with hadronic final states.
The dominant background is from the process $e^{+} e^{-} \rightarrow (\gamma) \mu^{+} \mu^{-}$, and the corresponding normalized numbers of surviving events are estimated and summarized in Table~\ref{CROSS}.
The background level, defined as the ratio of the number of the background events to that of the signal, varies from 0.5\% to 60\% depending on the c.m.\ energy. It is worth noting that no peaking background is found in the signal region.
The number of signal events is determined by subtracting the expected number of background events from the event yield in data.

\section{Cross section and form factor}
\subsection{Signal yields}
\label{SecXS}
\begin{figure}[h]
	\centering
	\includegraphics[scale=0.42]{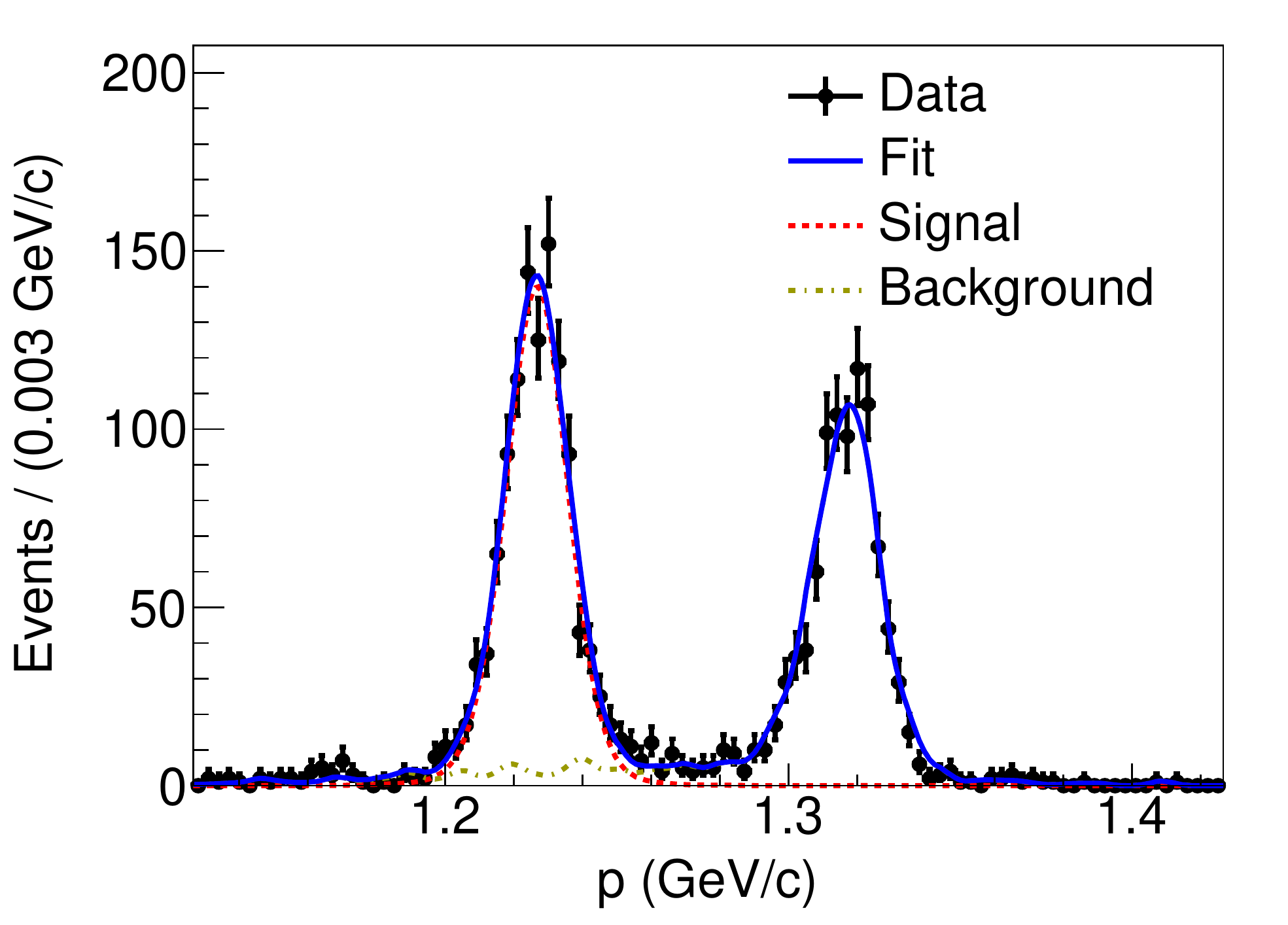}
	\caption {(color online) Momentum spectrum of the positive charged track for the data sample at $\sqrt{s} = 2.6444$~GeV. The solid line represents the total fit function, while the red and green dashed lines are the signal (main part of left peak) and the $(\gamma)\mu^{+}\mu^{-}$ background (right peak and its tail), respectively.}
	\label{pspecfit}
\end{figure}

\indent
The signal yields are determined by an unbinned maximum likelihood fit to the momentum distribution of the positive charged track of selected events, with the additional requirement on the momentum of the negative track to be in the interval $(p_{\rm exp}-3\sigma_{p}, p_{\rm exp}+3\sigma_{p})$.
In the fit, the signal shape is described by that of signal MC simulation convolved with a Gaussian function, which takes account the resolution difference between data and MC simulation.
Since the background is dominated by the process $e^{+} e^{-} \rightarrow (\gamma) \mu^{+} \mu^{-}$, the corresponding shape in the fit is described with the MC shape of the $e^{+} e^{-} \rightarrow (\gamma) \mu^{+} \mu^{-}$ process convolved with another Gaussian function. The distribution and the corresponding fit curve of the momentum of the positive charged track for the data sample at $\sqrt{s}=2.6444$ GeV is shown in Fig.~\ref{pspecfit}.

\subsection{Efficiency and correction factor}
\indent
The Born cross section is calculated from

\begin{equation}
\sigma^{B} = \frac{N_{\rm sig}}{\mathcal{L}\cdot\epsilon\cdot(1+\delta)},
\label{CX}
\end{equation}

\noindent
where $N_{\rm sig}$ is the number of signal events,
$\mathcal{L}$ the integrated luminosity measured with the method described in Ref.~\cite{GaoZhen},
$\epsilon$ the detection efficiency and
$1+\delta$ is the correction factor due to ISR and vacuum polarization (VP).

Both $\epsilon$ and $1+\delta$ are obtained from MC simulations of the signal reaction at the individual c.m.~energies.
In the \textsc{conexc} generator \cite{CONEXC}, the cross section for the ISR process ($\sigma_{e^{+}e^{-}\rightarrow \gamma X}$) is parameterized using

\begin{equation}
\sigma_{e^{+}e^{-}\rightarrow \gamma X} = \int d\sqrt{s'}\frac{2\sqrt{s'}}{s}W(s,x)\frac{\sigma^{B}(s')}{[1-\Pi(s')]^2},
\end{equation}

\noindent
where $\sqrt{s'}$ is the effective c.m.~energy of the final state with $s' = s(1-x)$, $x$ depends on the energy of the radiated photon according to $x = 2E_{\gamma}/\sqrt{s}$, $W(s, x)$ is the radiator function and
$\Pi(s')$ describes the VP effect. The latter includes contributions from leptons and quarks.
The detection efficiency $\epsilon$ and the radiative correction factor $1+\delta$ depend on the input cross section, and can only be extracted by an iterative procedure,
in which the line shape of the cross section obtained from {BaBar}~\cite{BABARKK} is used as the initial cross section, and the updated Born cross section is obtained according to the simulation.
We repeat the procedure until the measured Born cross section does not change by more than 0.5\%.

For the data samples with c.m.~energies larger than 3~GeV, near the $J/\psi$ resonance, the interference between the resonant process $J/\psi~\rightarrow~K^{+}~K^{-}$ and the continuum process $e^{+} e^{-} \rightarrow K^{+} K^{-}$ occurs.
To account for the interference, another data sample collected in the vicinity of the $J/\psi$ resonance is used to determine the correction factor for the interference.
A function including the amplitudes of the $J/\psi$ decay and the continuum process is used to fit the line shape of the measured cross section, and the ratio of continuum contribution to the total cross section is taken as the correction factor.
The resulting Born cross sections and related variables are summarized in Table~\ref{CROSS}.

\subsection{Line shape of \begin{boldmath}$e^{+} e^{-} \rightarrow K^{+} K^{-}$\end{boldmath}}

\begin{figure}[htbp]
	\centering
	\includegraphics[scale=0.42]{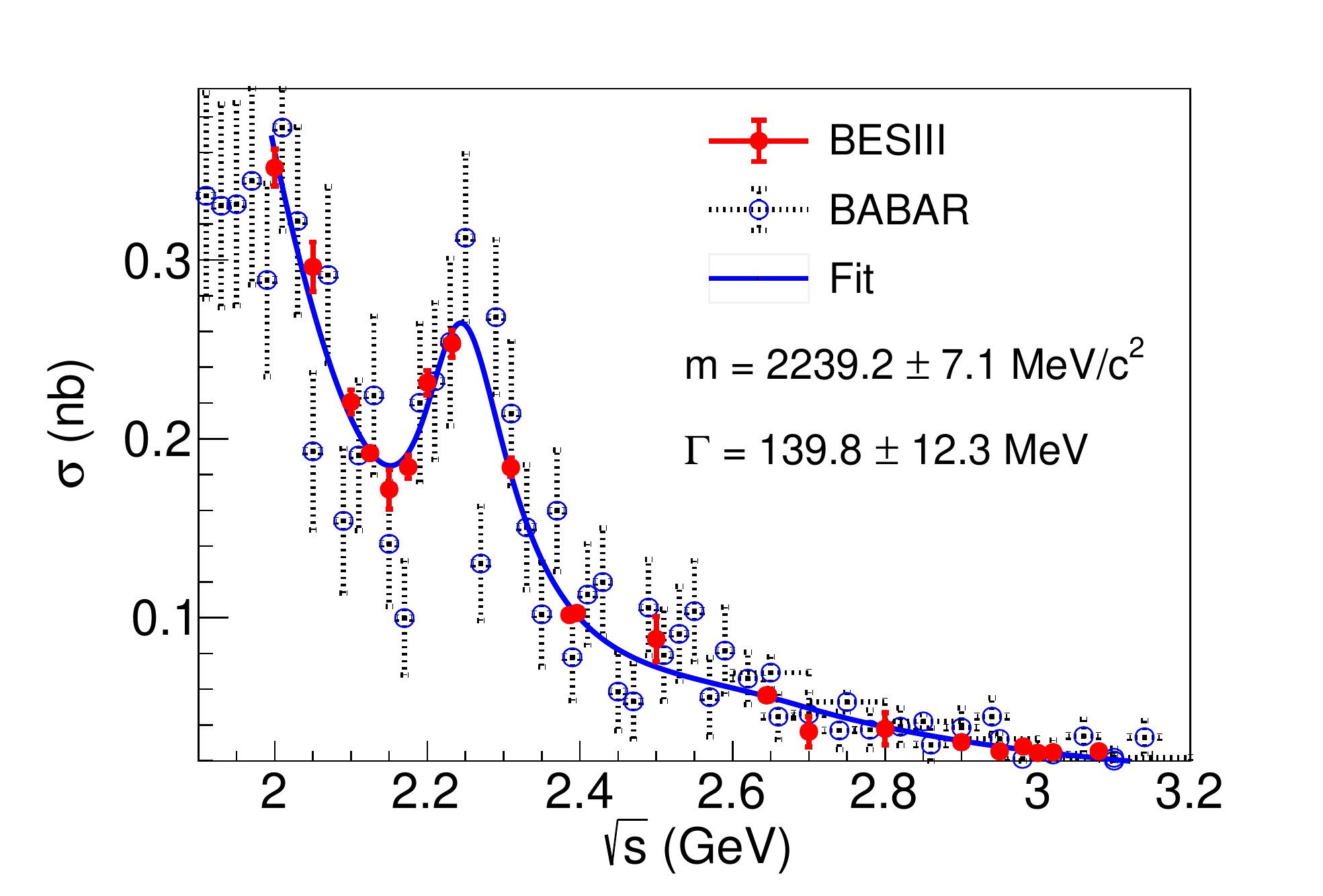}
\caption {(color online) Born cross section of the $e^{+} e^{-} \rightarrow K^{+} K^{-}$ process.  
Open black dots and filled triangles with error bars are the results of {BaBar} \cite{BABARKK, BABARKK2}. 
Red solid dots show the results of BESIII (this work). 
The error bars include both statistical and systematic uncertainties. 
The fit shown is performed using the BESIII result using Eq.~(\ref{equ::xsmr}).}
\label{XSbes}
\end{figure}

\begin{figure}[htbp]
	\centering
	\includegraphics[scale=0.38]{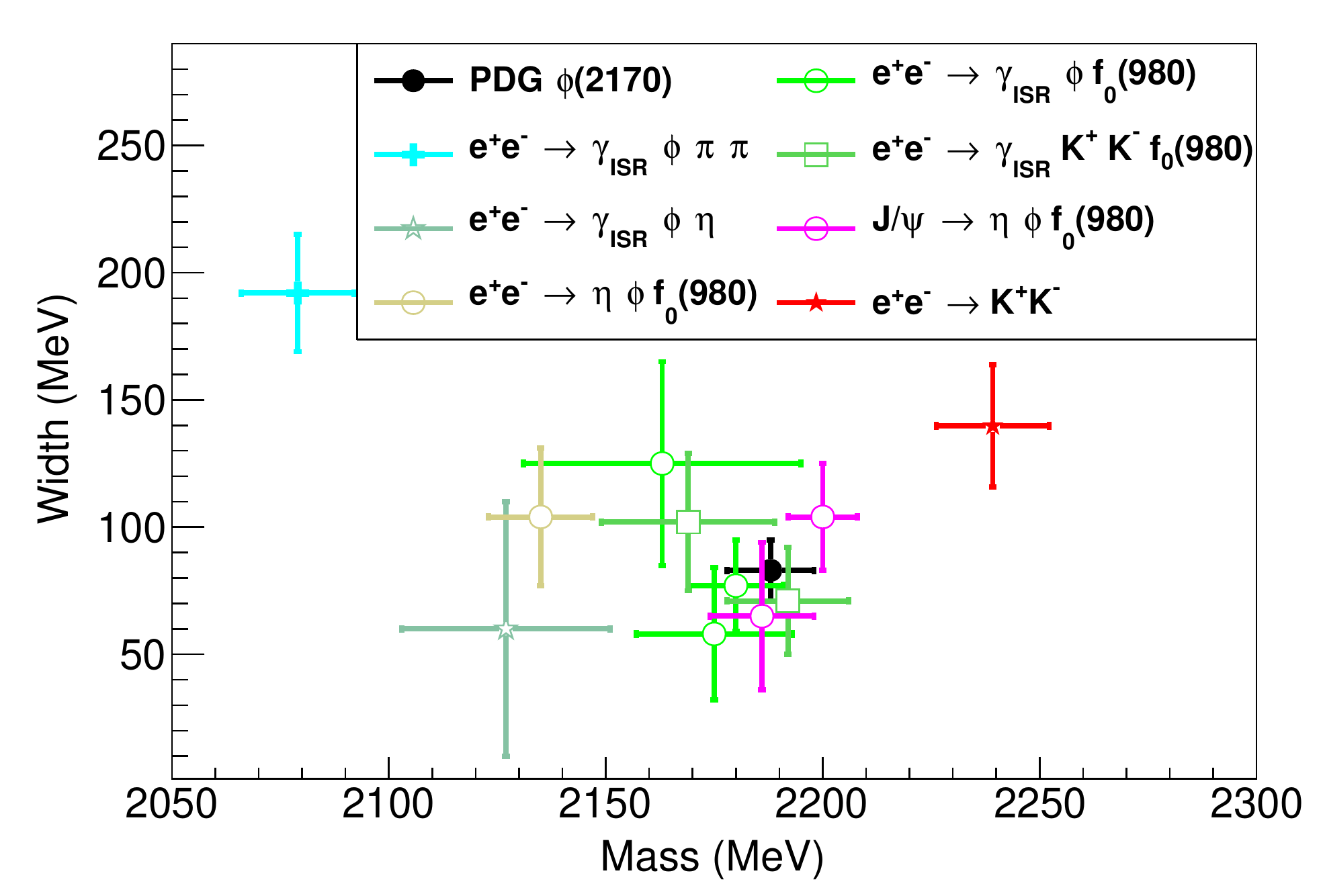}
\caption {(color online) Parameters of the $\phi(2170)$ state obtained from different processes and the resonance in the $e^{+}e^{-}\rightarrow~K^{+}K^{-}$ process. }
\label{cmpphi}
\end{figure}

The measured Born cross sections are shown in Fig.~\ref{XSbes}, where a clear structure is observed around 2.23~GeV. The cross sections are consistent with those of {BaBar}~\cite{BABARKK, BABARKK2}, and have better precision comparing to any previous measurement~\cite{DM2KK,CMD2KK, BABARKK,BABARKK2,eeSUM,BerKK,CLEOKK}.
A $\chi^2$ fit incorporating the correlated and uncorrelated uncertainties is performed to the measured cross section with the function

\begin{equation}
\label{equ::xsmr}
\sigma^{B} = |c_{R}\cdot BW_{R} + c_{con} \cdot s^{-\alpha} \cdot e^{i\cdot \theta1} + P \cdot e^{i\cdot \theta2} |^2,
\end{equation}

\noindent
where $c_{i}$ is the magnitude of component $i$, $R$ denotes the component for a structure around 2.23~GeV,
the term $s^{-\alpha}$ parameterizes the continuum process, $P$ is a polynomial function used to compensate unknown contributions,
$\theta_1$ and $\theta_2$ are the phases of the continuum and unknown components relative to the structure around 2.23~GeV, respectively.
$BW$ is a Breit-Wigner function for the structure around 2.23~GeV,  takes the form,

\begin{equation}
BW(s,m,\Gamma) = \frac{1}{m^2c^4-s-i\sqrt{s}\Gamma},
\end{equation}

\noindent
where $m$ and $\Gamma$ are the mass and width of the resonance, respectively.
In the fit, both statistical and systematic uncertainties are taken into account. 
Uncertainties from the ISR and the VP correction factor, the luminosity, and the tracking efficiency are assumed to be correlated across the whole range in $\sqrt{s}$, while the remaining uncertainties are treated to be uncorrelated.
The fit curve is shown in Fig.~\ref{XSbes}. The parameters of structure are determined to be $m =$~\massRes~and $\Gamma =$~\widthRes. 
To understand its nature, the result is compared with the parameters of $\phi(2170)$ state measured by previous experiments via various processes as shown in Fig. \ref{cmpphi}. The result differs from the world average parameters of the $\phi(2170)$ state by more than $3\sigma$ in mass and more than $2\sigma$ in width, and also differs from most individual experiments.

\input{cross_draft_v8.tex}

\subsection{The form factor}
\label{Sec::FF}
The electromagnetic form factor of the charged kaon can be extracted from the production cross section by assuming one-photon exchange \cite{BABARKK}:

\begin{equation}
|F_K|^2(s) = \frac{3s}{\pi \alpha(0)^2 \beta_{K}^{3}} \frac{\sigma^{D}}{C_{\rm FS}},
\label{EquFF}
\end{equation}

where

\begin{equation}
\sigma^{D} = \sigma^{B}\left(\frac{\alpha(s)}{\alpha(0)}\right)^2 
\label{DRESSCX}
\end{equation}

\noindent
is the dressed cross section,
$\alpha(s)$ the electromagnetic coupling constant,
$\beta_{K} = \sqrt{1-4m_{K}^2c^4/s}$ is the kaon velocity
and $C_{\rm FS}$ is the final-state radiative correction  for radiative effects \cite{FS1,FS2,FS3}.
The calculated form factors are listed in Table~\ref{CROSS}.

\begin{figure}[b]
	\centering
	\includegraphics[scale=0.42]{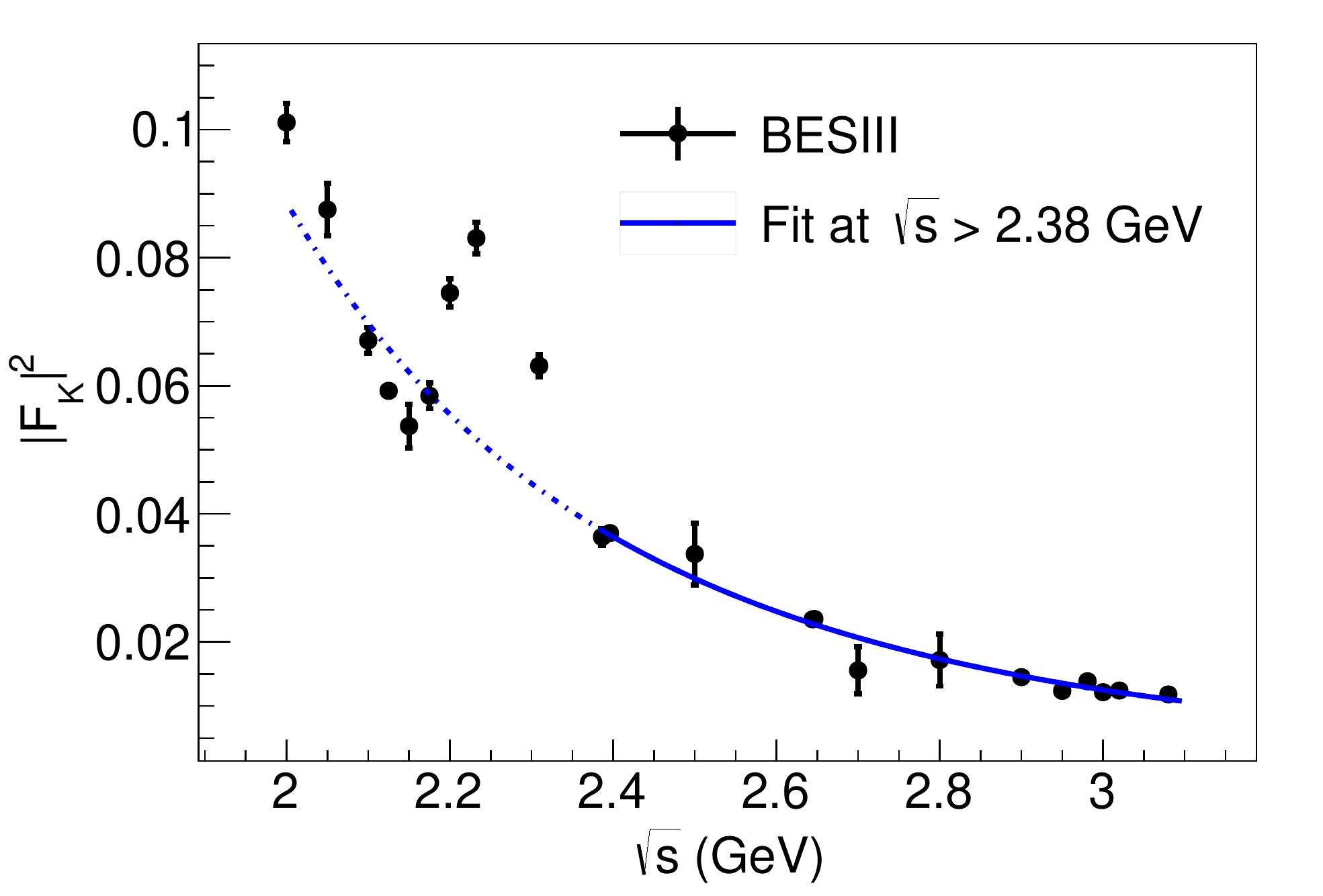}
	\caption {
	Distribution of $|F_{K}|^2$ for the process $e^{+} e^{-} \rightarrow K^{+} K^{-}$. Dots with error bar (in black) show the measured $|F_{K}|^2$. The solid line (in blue) is the fit result at $\sqrt{s}~>~2.38$ GeV, and the dotted line is the extrapolation of the fit towards smaller $\sqrt{s}$ to show the trend of the QCD prediction at lower energy.
	}
	\label{FFcmp}
\end{figure}

From pQCD, the form factor of a spin zero meson is predicted to be $F_{K} = 16 \pi \alpha_{s}(s){f_K^2}/{s}$ \cite{QCDFF}, where $\alpha_{s}(s)$ is the strong coupling constant and $f_K$ is the decay constant of the charged kaon.
 A $\chi^2$ fit incorporating the correlated and uncorrelated uncertainties to the $|F_{K}|^2$ distribution is performed with a function $A\alpha_{s}^{2}(s)/s^{n}$ for the data samples with c.m.~energy $\sqrt{s}>2.38~{\rm GeV}$ only, to avoid the influence of the structure around 2.23~GeV. The fit is shown in Fig.~\ref{FFcmp}, and yields the parameter $n$ to be $n = 1.94 \pm 0.09$, which is in agreement with the QCD
 prediction $n=2$.
At lower energies, the pQCD prediction is not valid, and no fit is performed in this analysis.

\section{Systematic uncertainty}

Several sources of systematic uncertainties, namely from detection efficiency, luminosity, ISR and VP correction factors, and the fit procedure for the signal extraction, are considered in the measurement of the Born cross section and the charged kaon form factors, as discussed in the following.

The sources of the uncertainty associated with the detection efficiency include tracking efficiency, selection criteria on the momentum of the negative charged tracks, $E/p$, $\cos\theta$ and the opening angle as well as the uncertainty due to the limited MC sample size.
The uncertainty in the tracking efficiency is studied with a control sample of $ e^{+}e^{-} \rightarrow K^{+}K^{-}\pi^{+}\pi^{-}$ by implementing the same strategy described in Ref.~\cite{TRKERR}. In this analysis, the kaons have momenta ranging from 0.85 to 1.45 GeV/$c$, and the transverse-momentum-weighted uncertainty of tracking efficiency is 1\% per track.
To study the uncertainties associated with the requirement on $p$, $E/p$ and opening angle criteria, we compared the distributions of corresponding variables between data and MC simulation, smeared the MC sample to match the data, and re-calculated the detection efficiency and cross section, individually. The resulting changes in the cross sections are taken as systematic uncertainties. 
The uncertainty due to the requirement on $\cos\theta$ is small and ignored in the analysis. 
The uncertainty related with MC statistics is estimated by
$\Delta_{MC} = \sqrt{(1-\epsilon)/{\epsilon}}/\sqrt{N}$, where $N$ is the number of signal MC events.
The integrated luminosities of the individual c.m.~energy points are measured using large-angle Bhabha scattering events, with an uncertainty of 0.9\% \cite{GaoZhen}, which is taken as the systematic uncertainty.
During the analysis, the cross section is measured by iterating until $(1+\delta)\epsilon$ converges, and the difference between the last two iterations is taken as the systematic uncertainty associated with the ISR and VP correction factors.
In this analysis, the signal yields are determined by a fit to the momentum spectrum of positive charged tracks. The uncertainties associated with the signal and background shapes, as well as the fit range are considered.
Uncertainties due to the choice of the signal and background shapes are estimated by changing signal and background functions to analytical Crystal Ball functions.
Uncertainties associated with the fit range are estimated by enlarging or shrinking the fit range by the momentum resolution.
The kaon form factors are extracted from the cross section and share the systematic uncertainties.
All systematic uncertainties of the cross section measurement and kaon form factor are summarized in Table~\ref{ERR}.\\

\input{uncertainty_draft_v9.tex}

The systematic uncertainties of the resonance parameters come from the absolute c.m.~energy measurement, the uncertainty of the measured cross section, and the fit procedure.
The uncertainty of the c.m.~energy from BEPCII is small and is found to be negligible in the determination of the resonance parameters.
The statistical and systematic uncertainties of the measured cross section has been considered in the fit of the cross section line shape, thus no further consideration in estimating the systematic uncertainties of resonance parameters is necessary.
The uncertainties associated with the fit procedure include those from the fit range and from the signal and background models.
The uncertainty from the fit range is investigated by excluding the first energy point $\sqrt{s} = 2.00$~GeV and last energy point $\sqrt{s} = 3.08$~GeV in the fit.  The changes with respect to the nominal result, $7.2$~MeV/$c^{2}$ for the mass and $20.2$~MeV for the width are taken as the systematic uncertainties.
To assess the systematic uncertainty associated with the signal model, a modified Breit-Wigner function, whose width is energy-dependent, is used in the fit, resulting in differences of $5.9$~MeV/$c^{2}$ and $1.7$~MeV for mass and width, respectively.
The uncertainty due to the function used to describe the contribution other than the signal structure is estimated by a fit combining {BaBar} and BESIII data.  The changes are found to be 6.4 MeV/$c^{2}$ and 3.5 MeV for mass and width, respectively.
The overall systematic uncertainties are obtained by summing all independent uncertainties in quadrature; they are
$11.3$~MeV/$c^2$ for the mass and $20.6$~MeV for the width.

\section{Conclusion}
In summary, we have measured the Born cross section of $e^{+} e^{-} \rightarrow K^{+} K^{-}$ and the charged kaon form factor using data samples collected with the BESIII detector at 22 different c.m.~energies from 2.00 to 3.08 GeV. 
The measured cross sections are consistent with those of {BaBar} and are of the best precision compared to previous measurements.
A clear structure is observed in the line shape of the measured cross section, and a fit yields a mass of \massResys{} and a width of \widthResys{} for this structure, where the first uncertainties are statistical and the second ones are systematic.
The extracted electromagnetic form factor of the charged kaon is fitted at c.m.~energies above $2.38$~GeV, and shows consistence with the pQCD prediction of $|F_{K}|$ decreasing with $1/s$.

From the Particle Data Group~\cite{Tanabashi:2018oca}, possible candidates for the observed structure may be the $\rho(2150)$ or $\phi(2170)$ meson.
Although the measured parameters agree within $2\sigma$ 
with those from some individual experiments, the results obtained in this 
paper differ from the world average parameters of $\rho(2150)$ and $\phi(2170)$ by more than $3\sigma$ in mass and more than $2\sigma$ in width. 
For the $\phi(2170)$ case, the result deviates from almost all individual measurements in the $e^{+}e^{-}$ annihilation process, disfavoring the reaction $e^+e^- \to \phi(2170) \to K^+K^-$.
Thus, the coupling of $\phi(2170)$ to $K^{+}K^{-}$ is also disfavored, and this may help to veto the model that treats $\phi(2170)$ as a $2^{3}D_{1}$ state of the $s\bar{s}$ system \cite{Ding:2007pc}. 
For the $\rho(2150)$ case, the result is consistent with the measurement in the process $e^{+}e^{-}~\rightarrow~\gamma \pi^{+}\pi^{-}$ \cite{Lees:2012cj}, which is not used in the world average.
Nevertheless, the nature of the resonance calls for further more detailed studies, like a combined analysis with other final states, or a partial wave analysis.

\input{acknowledgement}

\end{document}

%% file: authors.tex
\author{
\begin{small}
\begin{center}
M.~Ablikim$^{1}$, M.~N.~Achasov$^{10,d}$, S. ~Ahmed$^{15}$, M.~Albrecht$^{4}$, M.~Alekseev$^{56A,56C}$, A.~Amoroso$^{56A,56C}$, F.~F.~An$^{1}$, Q.~An$^{53,43}$, J.~Z.~Bai$^{1}$, Y.~Bai$^{42}$
% ...
\\
\vspace{0.2cm}
(BESIII Collaboration)\\
\vspace{0.2cm} {\it
$^{1}$ Institute of High Energy Physics, Beijing 100049, People's Republic of China\\
% ...
$^{4}$ Bochum Ruhr-University, D-44780 Bochum, Germany\\
% ...
$^{10}$ G.I. Budker Institute of Nuclear Physics SB RAS (BINP), Novosibirsk 630090, Russia\\
% ...
$^{15}$ Helmholtz Institute Mainz, Johann-Joachim-Becher-Weg 45, D-55099 Mainz, Germany\\
% ...
$^{42}$ Southeast University, Nanjing 211100, People's Republic of China\\
$^{43}$ State Key Laboratory of Particle Detection and Electronics, Beijing 100049, Hefei 230026, People's Republic of China\\
% ...
$^{53}$ University of Science and Technology of China, Hefei 230026, People's Republic of China\\
% ...
$^{56}$ (A)University of Turin, I-10125, Turin, Italy; (B)University of Eastern Piedmont, I-15121, Alessandria, Italy; (C)INFN, I-10125, Turin, Italy\\
% ...
$^{d}$ Also at the Novosibirsk State University, Novosibirsk, 630090, Russia\\
% ...
}\end{center}
\vspace{0.4cm}
\end{small}
}

\noaffiliation{}

%% file: cross_draft_v8.tex
{
\setlength{\tabcolsep}{7.5pt}
\begin{table*}[htbp] 
\begin{center}
\caption{\small
Cross sections of the $e^{+}e^{-} \rightarrow K^{+} K^{-}$ process and form factors of kaon. 
$N_{\rm sig}$ is the number of signal events, excluding the number of survived $\mu^{+}\mu^{-}$ events $N_{\mu\mu}^{\rm MC}$ 
in the signal region estimated from MC simulation, along 
with detection efficiency $\epsilon$, radiative and VP correction 
factor $1+\delta$, and luminosity $\mathcal{L}$.
$\sigma^{B}$ is the measured Born cross section, from which the form factor $F_{K}$ is extracted.
The first uncertainties are statistical and the second ones
systematic. Uncertainties on the form factor are propagated from those
on the cross sections.}
\label{CROSS}
	\begin{tabular*}{\textwidth}{ c c c   r@{.}l   r@{\ $\pm$\ }l     r@{.}l    r@{\ $\pm$\ }c@{\ $\pm$\ }l    c}
	\hline
	\hline
	$\sqrt{s}$ (GeV) & $\epsilon$ & $1+\delta$  & \multicolumn{2}{c}{$\mathcal{L}$} (pb$^{-1})$ & \multicolumn{2}{c}{$N_{\rm sig}$} & \multicolumn{2}{c}{$N_{\mu\mu}^{\rm MC}$} & \multicolumn{3}{c}{$\sigma^{B}$ (pb)} & $|F_{K}|^{2}$ \\
	\hline
	2.0000	& 0.1927	& 2.717     &  10&1  & 1853.8 & 43.3  &  9&0    &  351.5 	& 8.2	 &9.0 	&  0.1021  $\pm$ 0.0024 $\pm$ 0.0026 \\%
	2.0500	& 0.1853	& 2.864     &   3&34 & 525.4  & 23.2  &  2&6    &  296.1 	&13.1	 &7.5 	&  0.0878  $\pm$ 0.0039 $\pm$ 0.0022 \\%
	2.1000	& 0.1591	& 3.368     &  12&2  & 1438.0 & 38.3  & 14&9    &  220.6 	& 5.9	 &5.5 	&  0.0666  $\pm$ 0.0018 $\pm$ 0.0017 \\%
	2.1250	& 0.1453	& 3.704     & 109&   & 11209.5& 106.9 &125&3    &  192.0 	& 1.8	 &4.7 	&  0.0593  $\pm$ 0.0006 $\pm$ 0.0015 \\%
	2.1500	& 0.1346	& 3.987     &   2&84 & 261.7  & 16.3  &  2&6    &  171.7 	&10.7	 &4.2 	&  0.0539  $\pm$ 0.0034 $\pm$ 0.0013 \\%
	2.1750	& 0.1521	& 3.521     &  10&6  & 1048.1 & 32.7  & 12&1    &  184.2 	& 5.7	 &4.6 	&  0.0590  $\pm$ 0.0018 $\pm$ 0.0015 \\%
	2.2000	& 0.1802	& 2.986     &  13&7  & 1706.0 & 41.7  & 24&4    &  231.4 	& 5.7	 &6.0 	&  0.0744  $\pm$ 0.0018 $\pm$ 0.0019 \\%
	2.2324	& 0.2011	& 2.707     &  11&9  & 1634.2 & 40.8  & 17&1    &  253.2 	& 6.3	 &6.4 	&  0.0843  $\pm$ 0.0021 $\pm$ 0.0021 \\%
	2.3094	& 0.1697	& 3.255     &  21&1  & 2143.3 & 46.9  & 34&3    &  184.0 	& 4.0	 &4.8 	&  0.0635  $\pm$ 0.0014 $\pm$ 0.0017 \\%
	2.3864	& 0.1222	& 4.557     &  22&6  & 1274.9 & 36.4  & 40&0    &  101.5 	& 2.9	 &2.8 	&  0.0367  $\pm$ 0.0010 $\pm$ 0.0010 \\%
	2.3960	& 0.1189	& 4.702     &  66&9  & 3837.3 & 63.2  &148&0    &  102.6 	& 1.7	 &2.9 	&  0.0371  $\pm$ 0.0006 $\pm$ 0.0010 \\%
	2.5000	& 0.1005	& 5.616     &   1&10 & 54.6   & 7.6   &  2&1    &  88.1  	&12.2	 &2.8 	&  0.0341  $\pm$ 0.0047 $\pm$ 0.0011 \\%
	2.6444	& 0.0909	& 6.289     &  33&7  & 1091.9 & 34.7  &110&4    &  56.6  	& 1.8	 &2.3 	&  0.0237  $\pm$ 0.0008 $\pm$ 0.0010 \\%
	2.6464	& 0.0902	& 6.300     &  34&0  & 1095.3 & 34.9  &100&0    &  56.7  	& 1.8	 &1.8 	&  0.0240  $\pm$ 0.0008 $\pm$ 0.0008 \\%
	2.7000	& 0.0873	& 6.580     &   1&03 & 21.6   & 5.0   &  3&4    &  36.3  	& 8.4	 &1.3 	&  0.0158  $\pm$ 0.0037 $\pm$ 0.0006 \\%
	2.8000	& 0.0804	& 7.159     &   1&01 & 22.1   & 5.1   &  4&1    &  37.9  	& 8.8	 &1.7 	&  0.0173  $\pm$ 0.0040 $\pm$ 0.0007 \\%
	2.9000	& 0.0738	& 7.837     & 105&   & 1847.8 & 48.1  &496&0    &  30.4  	& 0.8	 &1.5 	&  0.0145  $\pm$ 0.0004 $\pm$ 0.0007 \\%
	2.9500	& 0.0702	& 8.217     &  15&9  & 232.9  & 17.3  & 87&0    &  25.3  	& 1.9	 &1.4 	&  0.0125  $\pm$ 0.0009 $\pm$ 0.0007 \\%
	2.9810	& 0.0683	& 8.466     &  16&1  & 260.6  & 15.1  & 87&2    &  28.0  	& 1.6	 &1.6 	&  0.0139  $\pm$ 0.0008 $\pm$ 0.0008 \\%
	3.0000	& 0.0667	& 8.622     &  15&9  & 215.5  & 16.9  & 89&8    &  24.4  	& 1.8	 &1.5 	&  0.0122  $\pm$ 0.0009 $\pm$ 0.0007 \\%
	3.0200	& 0.0656	& 8.791     &  17&3  & 235.9  & 18.2  & 99&3    &  24.8  	& 1.8	 &1.5 	&  0.0124  $\pm$ 0.0009 $\pm$ 0.0008 \\%
	3.0800	& 0.0564	& 9.266     & 126&   &1335.6  & 44.0  &863&5    &  25.3  	& 0.7	 &2.2 	&  0.0118  $\pm$ 0.0003 $\pm$ 0.0010 \\%
	\hline
	\hline
	\end{tabular*}
\end{center}
\end{table*}
}

%% file: uncertainty_draft_v9.tex
{
\setlength{\tabcolsep}{8.6pt}
\begin{table*}[htbp]
\caption{\small Summary of relative systematic uncertainties (in \%) associated with the luminosity ($\mathcal{L}$), the detection efficiency ($\epsilon$), the initial state radiation and the vacuum polarization correction factor ($1+\delta$), the momentum of the negative charged tracks ($p$),  the ratio of deposited energy and momentum ($E/p$), the opening angle (Angle), the tracking efficiency (Tracking), fit range (Fit), signal and background shapes (Sig.shape and Bck.shape).
 in the measurement of the Born cross section of the $e^{+}e^{-} \rightarrow K^{+} K^{-}$ process and charged kaon form factor. 
The total uncertainty is obtained by summing the individual
contributions in quadrature, noting that the uncertainties are also
considered in the correction of the $J/\psi$ contribution for energies higher than 3 GeV.}
\label{ERR}
\centering
	\begin{tabular}{c c c c c   c c c c c   c c}
	\hline
	\hline
	$\sqrt{s}$ (GeV)	&$\mathcal{L}$	&$\epsilon$  	&$1+\delta$  	&$p$         	&$E/p$       	&Angle	&Tracking	&Fit 	&Sig. shape    	&Bck. shape 	&Total      \\ 
	\hline
	2.0000	&0.9	&0.2	&0.2	&0.7	&0.6	&0.8	&2.0	&0.0	&0.2	&0.4	&2.5 \\
	2.0500	&0.9	&0.2	&0.1	&$0.1$	&0.7	&0.7	&2.0	&0.7	&0.2	&0.4	&2.5 \\
	2.1000	&0.9	&0.2	&0.3	&0.2	&0.5	&0.8	&2.0	&0.1	&0.2	&0.4	&2.5 \\
	2.1250	&0.8	&0.2	&0.3	&$0.1$	&0.6	&0.7	&2.0	&0.3	&0.2	&0.4	&2.4 \\
	2.1500	&0.9	&0.3	&0.5	&$0.1$	&0.6	&0.7	&2.0	&0.1	&0.3	&0.4	&2.5 \\
	2.1750	&0.9	&0.2	&0.3	&0.3	&0.6	&0.7	&2.0	&0.1	&0.4	&0.4	&2.5 \\
	2.2000	&0.9	&0.2	&0.3	&0.4	&0.6	&0.8	&2.0	&0.5	&0.4	&0.4	&2.6 \\
	2.2324	&0.9	&0.2	&0.5	&0.1	&0.5	&0.8	&2.0	&0.2	&0.5	&0.4	&2.5 \\
	2.3094	&0.9	&0.2	&0.2	&$0.1$	&0.6	&0.7	&2.0	&0.6	&0.7	&0.5	&2.6 \\
	2.3864	&0.9	&0.3	&0.4	&0.2	&0.4	&0.9	&2.0	&0.5	&1.0	&0.5	&2.7 \\
	2.3960	&0.9	&0.3	&0.4	&0.3	&0.4	&1.0	&2.0	&0.4	&1.0	&0.6	&2.8 \\ 
	2.5000	&0.9	&0.3	&0.2	&1.4	&0.6	&0.8	&2.0	&0.3	&1.3	&0.6	&3.2 \\
	2.6444	&0.9	&0.3	&0.3	&0.4	&0.6	&0.9	&2.0	&2.7	&1.7	&0.7	&4.1 \\
	2.6464	&0.9	&0.3	&0.3	&0.5	&0.6	&0.8	&2.0	&0.8	&1.7	&0.8	&3.2 \\
	2.7000	&0.9	&0.3	&0.3	&0.5	&0.4	&0.9	&2.0	&1.0	&2.0	&1.2	&3.6 \\
	2.8000	&0.9	&0.3	&0.3	&0.5	&0.7	&1.3	&2.0	&1.0	&2.5	&2.1	&4.4 \\
	2.9000	&0.9	&0.4	&0.3	&0.1	&0.4	&0.8	&2.0	&1.1	&3.0	&3.0	&5.0 \\
	2.9500	&0.9	&0.4	&0.3	&0.1	&0.4	&0.9	&2.0	&0.3	&3.3	&3.5	&5.4 \\
	2.9810	&0.9	&0.4	&0.3	&0.5	&0.5	&1.2	&2.0	&0.2	&3.4	&3.8	&5.7 \\
	3.0000	&0.9	&0.4	&0.3	&1.6	&0.4	&0.9	&2.0	&0.7	&3.5	&3.9	&6.1 \\
	3.0200	&0.9	&0.4	&0.3	&1.1	&0.5	&0.9	&2.0	&0.7	&3.6	&4.1	&6.2 \\
	3.0800	&0.9	&0.4	&0.3	&1.1	&0.4	&1.0	&2.0	&0.8	&3.9	&4.6	&8.9 \\ 
	\hline
	\hline
	\end{tabular}

\end{table*}
}

%% file: acknowledgement.tex
\section{Acknowledgement}
The BESIII collaboration thanks the staff of BEPCII, the IHEP computing center and the supercomputing center of USTC for their strong support. 
This work is supported in part by National Key Basic Research Program of China under Contract No. 2015CB856700; National Natural Science Foundation of China (NSFC) under Contracts Nos.~11235011, 11275189, 11322544, 11335008, 11375170, 11425524, 11475164, 11475169, 11605196, 11605198, 11625523, 11635010; 
the Chinese Academy of Sciences (CAS) Large-Scale Scientific Facility Program; the CAS Center for Excellence in Particle Physics (CCEPP); 
Joint Large-Scale Scientific Facility Funds of the NSFC and CAS under Contracts Nos.~U1332201, U1532102, U1532257, U1532258, 
CAS under Contracts Nos. KJCX2-YW-N29, KJCX2-YW-N45, QYZDJ-SSW-SLH003; 100 Talents Program of CAS; National 1000 Talents Program of China; INPAC and Shanghai Key Laboratory for Particle Physics and Cosmology; German Research Foundation DFG under Contracts Nos. Collaborative Research Center CRC 1044, FOR 2359; Istituto Nazionale di Fisica Nucleare, Italy; Koninklijke Nederlandse Akademie van Wetenschappen (KNAW) under Contract No. 530-4CDP03; Ministry of Development of Turkey under Contract No. DPT2006K-120470; National Science and Technology fund; The Swedish Research Council; U. S. Department of Energy under Contracts Nos. DE-FG02-05ER41374, DE-SC-0010118, DE-SC-0010504, DE-SC-0012069; University of Groningen (RuG) and the Helmholtzzentrum fuer Schwerionenforschung GmbH (GSI), Darmstadt; WCU Program of National Research Foundation of Korea under Contract No. R32-2008-000-10155-0.